# Definition and evolution of quantum cellular automata with two qubits per cell

Ioannis G. Karafyllidis

*Democritus University of Thrace*
*Department of Electrical and Computer Engineering*
*671 00 Xanthi, GREECE*
*email: ykar@ee.duth.gr*
*URL: http://vlsi.ee.duth.gr/~ykar*

**Abstract**

Studies of quantum computer implementations suggest cellular quantum computer architectures. These architectures can simulate the evolution of quantum cellular automata, which can possibly simulate both quantum and classical physical systems and processes. It is however known that except for the trivial case, unitary evolution of one-dimensional homogeneous quantum cellular automata with one qubit per cell is not possible. Quantum cellular automata that comprise two qubits per cell are defined and their evolution is studied using a quantum computer simulator. The evolution is unitary and its linearity manifests itself as a periodic structure in the probability distribution patterns.

# I. INTRODUCTION

Cellular Automata (CAs) were proposed by von Neumann as a model of self-replicating systems [1]. One-dimensional classical CAs with one bit per cell a were studied by Wolfram [2]. Since then, the CA model has been successfully used for the simulation of physical systems and processes and served as basis for parallel computer processor architectures [3 - 5]. CAs as mathematical objects can simulate complex physical phenomena and can be simulated exactly by digital computers, because the topology and evolution of the simulated object is reproduced in the simulating device [6]. Since CAs are a successful model for classical systems and an efficient basis for classical computer architectures, two questions follow naturally: Can CAs serve as model for quantum systems? Can CAs serve as quantum computer architecture?

Feynman examined the possibility of using CAs both as models of quantum systems and as quantum computer architecture in 1982 [7]. Grössing and Zeilinger developed a CA model for quantum systems in which probability is conserved and the evolution is linear, but the criterion of unitary evolution is relaxed by introducing terms of the form $i\delta$ in the evolution rule. The evolution is unitary only for $\delta \rightarrow 0$ [8]. Meyer proposed quantum lattice gas automata as model for the simulation of physical processes [9]. CAs developed to model quantum systems are referred to as quantum cellular automata (QCAs).

All the above studies revealed that, except for the trivial case, unitary evolution of one-dimensional QCAs is impossible. This is known as the "no-go lemma", which states that in one dimension there exist no non-trivial homogeneous, local, linear QCA [10]. The no-go lemma stems from the non-cloning theorem and presents a major obstacle for the construction of cellular quantum computer architectures. On the other hand, studies of quantum computer implementations suggest that cellular architectures are the natural quantum computer architectures [11, 12]. This paper will contribute to the resolution of this contrast by introducing an alternative QCA structure not affected by the no-go lemma, because each cell comprises two qubits one of which never interacts with neighboring qubits. This QCA has linear unitary evolution and can possibly serve as quantum computer architecture.

In this work QCAs with two qubits per cell are defined. The first qubit is the *controlled* qubit (c-qubit) and the other is the *state* qubit (s-qubit). The evolution rules are unitary and are expressed as tensor product of unitary quantum gates. The rules are applied in two phases. In the first phase, the state of the c-qubit in each cell is changed according to the states of the s-qubits in the neighboring cells. In the second phase a two-input quantum gate or a combination of quantum gates is applied to the c-qubit and the s-qubit in each cell allowing thus the information flow in the QCA lattice and simultaneous change of all s-qubit states. The evolution of QCAs with two qubits per cell is studied using a quantum computer simulator developed by the author [13-14]. The linearity of QCA evolution manifests itself as a possible structure in the probability distribution patterns.

## II. DEFINITION OF QCAs WITH TWO QUBITS PER CELL

A CA is characterized by five basic properties: the number of spatial dimensions of its lattice, the possible states of the CA cells, the neighborhood of each CA cell (i.e. the set of the cells that influence its state), the evolution rule according to which the CA cell states evolve and the boundary conditions at the ends of the CA lattice. In the QCAs that we are about to define the same evolution rule applies to all cells at all times. The number of spatial dimensions of the QCA is one, i.e. it forms one-dimensional lattice.

Each QCA cell comprises two qubits the c-qubit and the s-qubit. The state of the $j^{th}$ QCA cell at computation step $t$ is written as: $\left| s_j^t c_j^t \right\rangle$. There are four base-states for each cell. During QCA evolution the state of the cell may be in any base-state superposition:

$$\left| s_j^t c_j^t \right\rangle = c_{0,j}^t \left| 00 \right\rangle + c_{1,j}^t \left| 01 \right\rangle + c_{2,j}^t \left| 10 \right\rangle + c_{3,j}^t \left| 11 \right\rangle \tag{1}$$

where $c_{n,j}^t$, $n = 0, 1, 2, 3$ are complex numbers and $t$ and $j$ are indices. The global state of the QCA at the computation step $t$, $|S^t\rangle$, is the tensor product of the states of its cells:

$$|S^t\rangle = |\cdots s_{j+1}^t c_{j+1}^t s_j^t c_j^t s_{j-1}^t c_{j-1}^t \cdots\rangle \qquad (2)$$

Three different neighborhood sets are considered for the QCAs defined here. The first neighborhood of the $j^{th}$ QCA cell comprises the same cell and the $j+1$ cell, the second the same cell and the $j-1$ cell and the third the same cell, the $j+1$ and the $j-1$ cells.

The cell states and therefore the global state of the QCA evolve according to the evolution rule. The QCAs defined here evolve according to the global unitary rule $R$. The evolution of the global state from computation step $t$ to computation step $t+1$ is given by:

$$|S^{t+1}\rangle = R|S^t\rangle \qquad (3)$$

The rule $R$ is applied in two phases. In the first phase, which is the interaction phase, the state of the c-qubit in each cell is changed according to the states of the s-qubits in the neighboring cells. This is achieved by applying quantum controlled gates. In the second phase, which is the evaluation phase, a two-input quantum gate or a combination of quantum gates is applied to the c-qubit and the s-qubit in each cell. Therefore, rule $R$ is given as a product of the unitary operators $R_I$ and $R_E$ which correspond to the interaction and evaluation phases:

$$|S^{t+1}\rangle = R|S^t\rangle = R_E R_I |S^t\rangle \qquad (4)$$

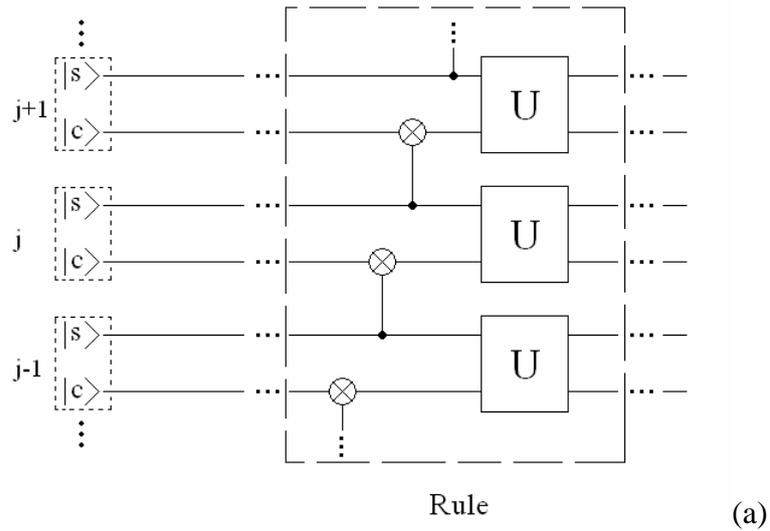

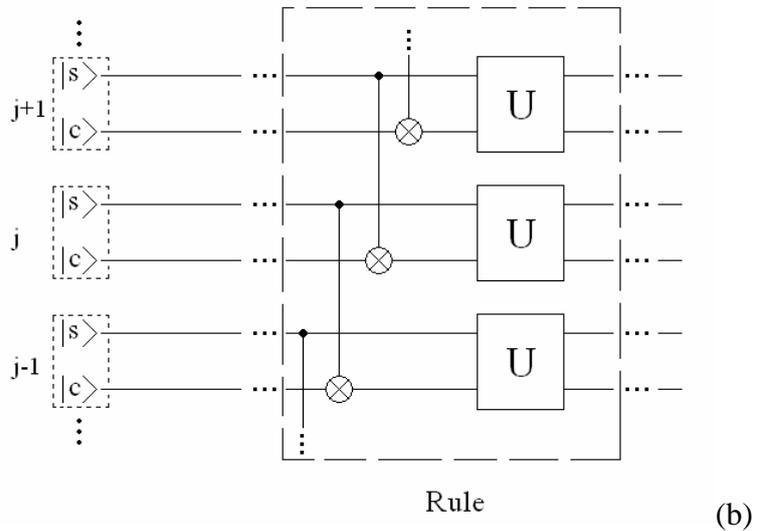

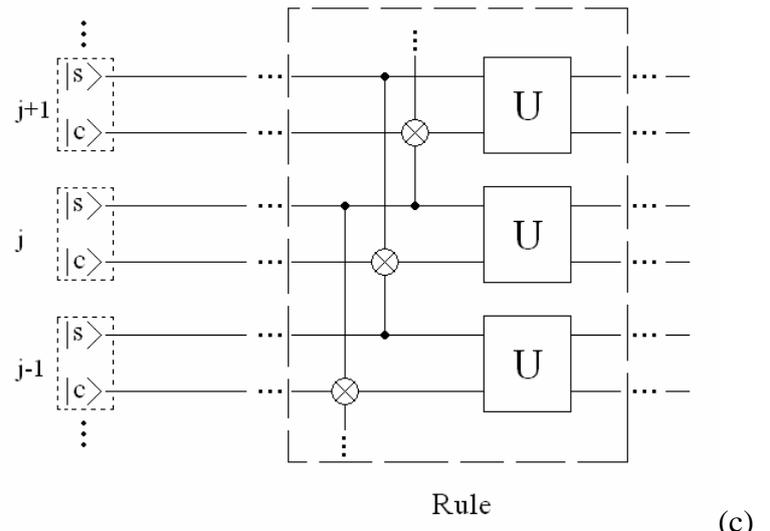

**FIG. 1.** *(a) The evolution rule for the case where the neighborhood comprises the $j^{th}$ and the $j+1$ cells. (b) The evolution rule for the case where the neighborhood comprises the $j^{th}$ and the $j-1$ cells. (c) The evolution rule for the case where the neighborhood comprises the $j+1$, the $j^{th}$ and the $j-1$ cells.*

FIGs. 1(a) and (b) show two QCA evolution rules as quantum circuits. In the case of FIG. 1(a) the neighborhood comprises the $j^{th}$ and the $j+1$ cells and in the case of FIG. 1(b) the $j^{th}$ and the $j-1$ cells. In the interaction phase a quantum Controlled-NOT (CN) gate is applied to the two cells of each neighborhood. In the case of FIG 1(a), the controlling qubit is the s-qubit of the $j^{th}$ cell and the controlled is the c-qubit of the $j+1$ cell. In the case of FIG 1(b) the controlling qubit is the s-qubit of the $j^{th}$ cell and the controlled qubit is the c-qubit of the $j-1$ cell. During this phase the states of all s-qubits remain unaltered and the states of all c-qubits change. The *CN* gates are applied to all neighborhoods and the operator $R_I$ is their tensor product:

$$R_I = \cdots \otimes CN \otimes CN \otimes CN \otimes \cdots \qquad (5)$$

In the evaluation phase a two-input quantum gate or a combination of quantum gates is applied to the c-qubit and the s-qubit in each cell. During this phase both c-qubit and s-qubit states may change. The two-input quantum gates or the combination of quantum gates applied in all cells are the same and the operator $R_E$ is their tensor product:

$$R_E = \cdots \otimes U \otimes U \otimes U \otimes \cdots \qquad (6)$$

where, *U* is the two-input quantum gate or a combination of quantum gates applied at each cell.

FIG. 1(c) shows the evolution rule for the case where the neighborhood comprises the $j+1$, the $j^{th}$ and the $j-1$ cells. In the interaction phase a quantum Controlled-Controlled-NOT (CCN) gate is applied to the three cells of each neighborhood. The controlling qubits are the s-qubits of the $j+1$ and $j-1$ cells and the controlled qubit is the c-qubit of the $j^{th}$ cell. During this phase only the states of the c-qubits change. In the evaluation phase a two-input quantum gate or a combination of quantum gates is applied to the c-qubit and the s-qubit in each cell. During this phase both c-qubit and s-qubit states may change. In this case operator $R_I$ is a tensor product of *CCN* gates. In all three neighborhoods the evolution rules are unitary operators since they are tensor products of quantum gates. In all one-dimensional evolution rules the nearest-neighbor connection scheme is used. This scheme stems from the one-dimensional classical CAs defined by Wolfram [2].

Constant and cyclic boundary conditions will be defined for the QCAs. In the case of constant boundary conditions a c-qubit at the end of the lattice is assumed to be controlled by a qubit which is constantly in states $|0\rangle$ or $|1\rangle$ and an s-qubit is assumed to control a qubit which does not participate in the QCA global state. In cyclic boundary conditions the *j-1* neighbor of the first cell is the last cell and the *j+1* neighbor of the last cell is the first cell.

## III. EVOLUTION OF QCAs WITH TWO QUBITS PER CELL

### A. The quantum computer simulator

The inputs to the simulator are the initial state of the qubits that form a quantum register (*QR*) and the quantum gates applied at each computation step. The initial state of the *QR* is entered as a one-column matrix, the elements of which are *0* and *1*, and the quantum gate configuration is entered as a two-dimensional matrix.

After the inputs are entered, the tensor product of the initial state of the *QR* is calculated and the simulator enters a loop. Iterations in this loop represent quantum computation steps. In each step the tensor product of the gate matrices applied in this step is calculated. Then, the new state of the *QR* is calculated as a product of the matrix that represents the quantum gates applied at this step and the matrix that represent the state of the *QR* at the previous step.

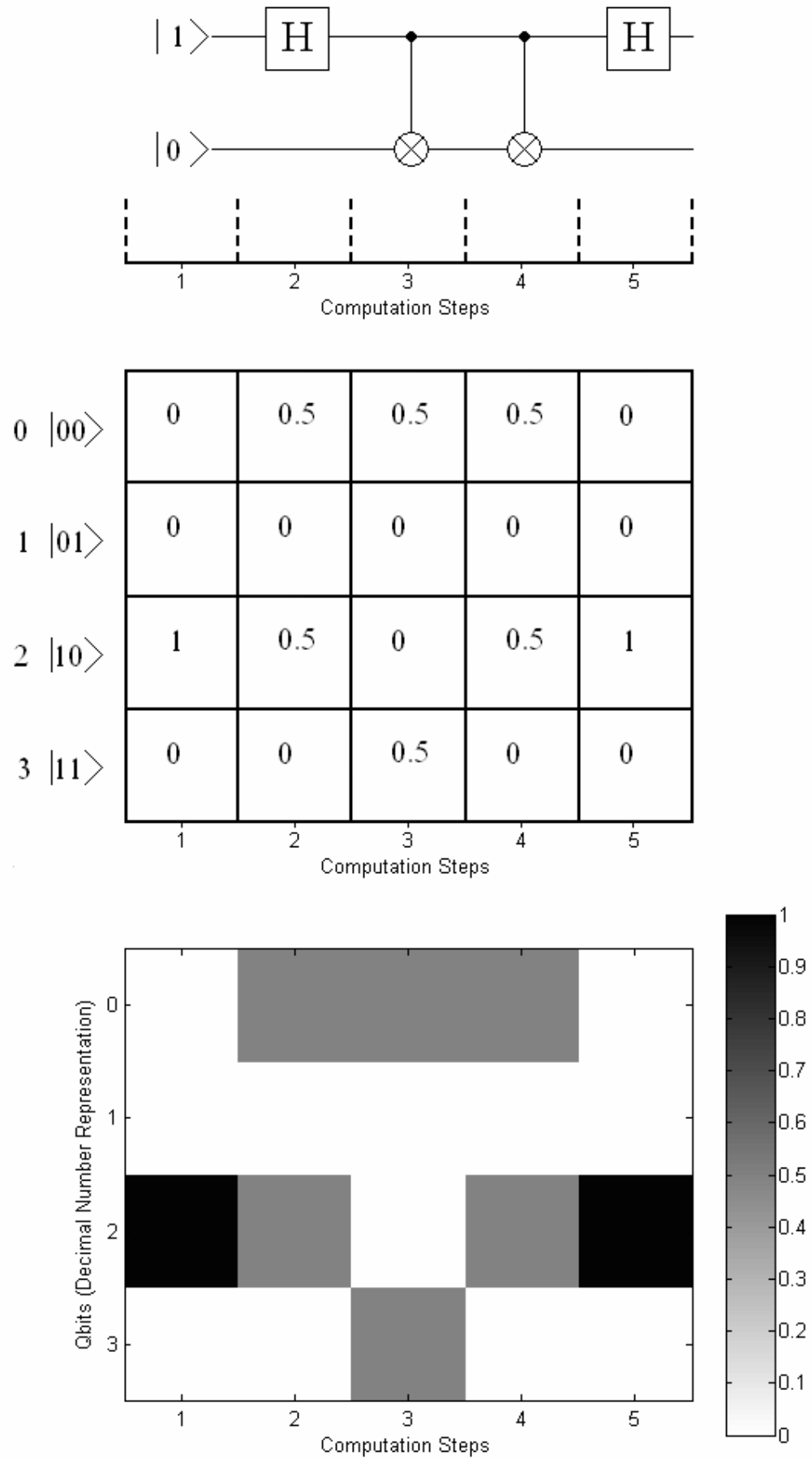

***FIG. 2.*** *A quantum computation and its simulation.*

After the end of the calculation, the simulator, apart from the matrices that represent the *QR* state at each quantum computation step, produces a graphical output that shows the probability of measuring each one of the possible *QR* states at each computation step. The probability is represented as a gray-scale image in which probability *1* is represented by black and probability *0* by white.

As an example, the quantum computation shown on top of FIG. 2 will be simulated. The columns of this matrix in the middle of FIG. 2 represent the computation steps and the rows the possible states of the quantum register, which are also written in decimal form. The first computation step is the initial state of the *QR*, which in this case is $|10\rangle$. A measurement of the *QR* at this step will give the state $|10\rangle$ with probability equal to *1*. All the matrix elements in the first column are zero except the element that corresponds to state *2* (in decimal), which is *1*. At the second step a Hadamard (H) gate is applied to the qubit with initial state $|1\rangle$. A measurement of the *QR* state at the end of this step will give states $|00\rangle$ and $|10\rangle$ with probability *0.5* and the matrix elements that correspond to states *0* and *2* (in decimal) are *0.5*. A *CN* gate is applied at the third step. A measurement at the end of this step will give states $|00\rangle$ and $|11\rangle$ (*0* and *3* in decimal) with probability *0.5* each, and so on. A graphical representation of this matrix is shown in the down part of FIG. 2. The *x-axis* represents the computation steps (matrix columns) and the *y-axis* the possible *QR* states in decimal form. Each matrix element corresponds to a rectangle and its value is represented as a gray-scale image in which probability *1* is represented by black and probability *0* by white.

## B. Evolution of QCAs

The evolution of QCAs with two states per cell were simulated using this quantum computer simulator. The evolution rules of FIG. 1 were applied many times so that the periodic structure in the probability patterns becomes apparent.

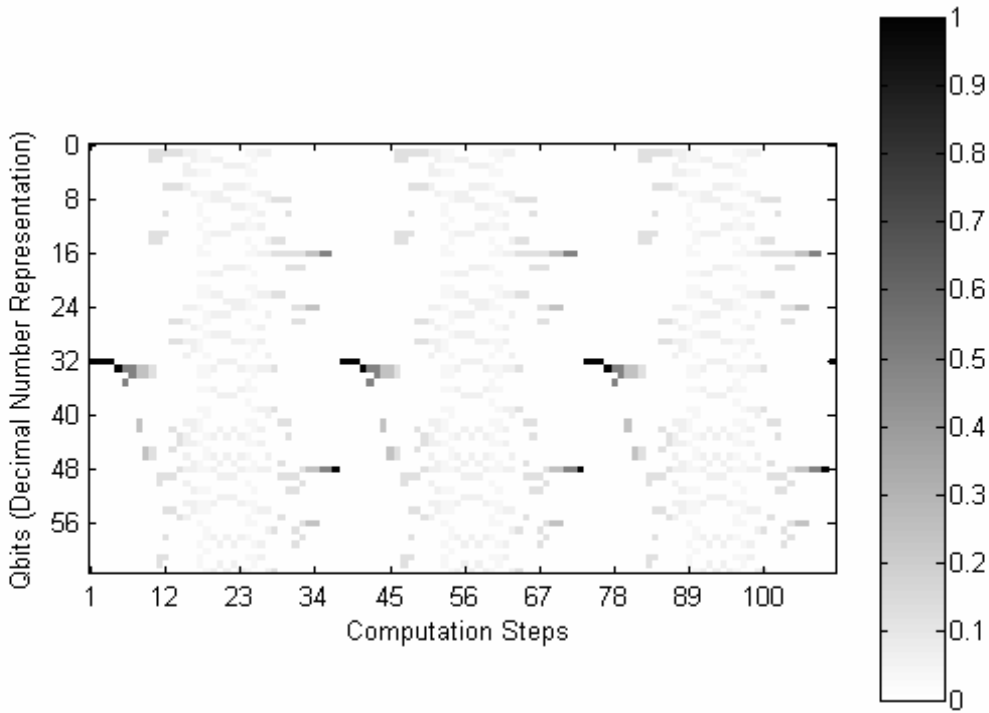

*FIG. 3. Evolution of a QCA with three cells. The rule is the one shown in FIG. 1(a). At evaluation phases the s-qubits and the c-qubits in each cell are entangled.*

FIG. 3 shows the simulated evolution of a QCA with three cells. The qubits are six and the number of base states is 64. On the *y-axis* the states are represented in decimal. The initial state of the QR is $|100000\rangle$, 32 in decimal and the evolution rule used is the one shown in FIG. 1(a). At evaluation phases a combination of quantum gates was used. An *H* gate was applied to the s-qubits in each cell followed by a *CN* gate with the s-qubit as controlling and the c-qubit as controlled qubit. That is, at the evaluation phases both qubits in each cell are entangled.

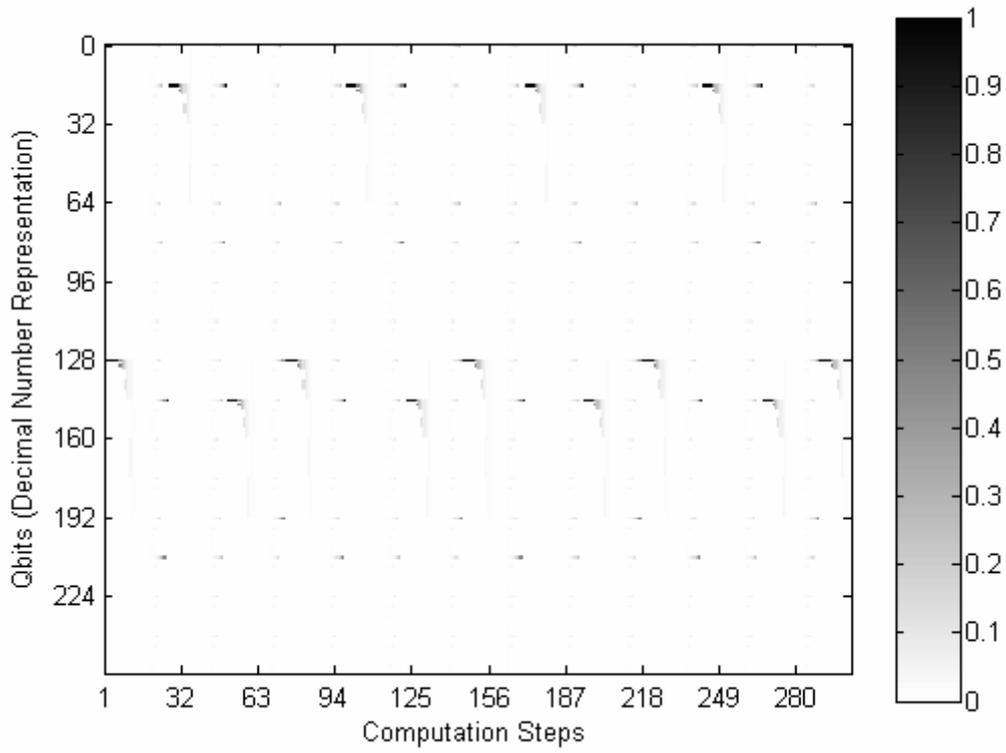

(a)

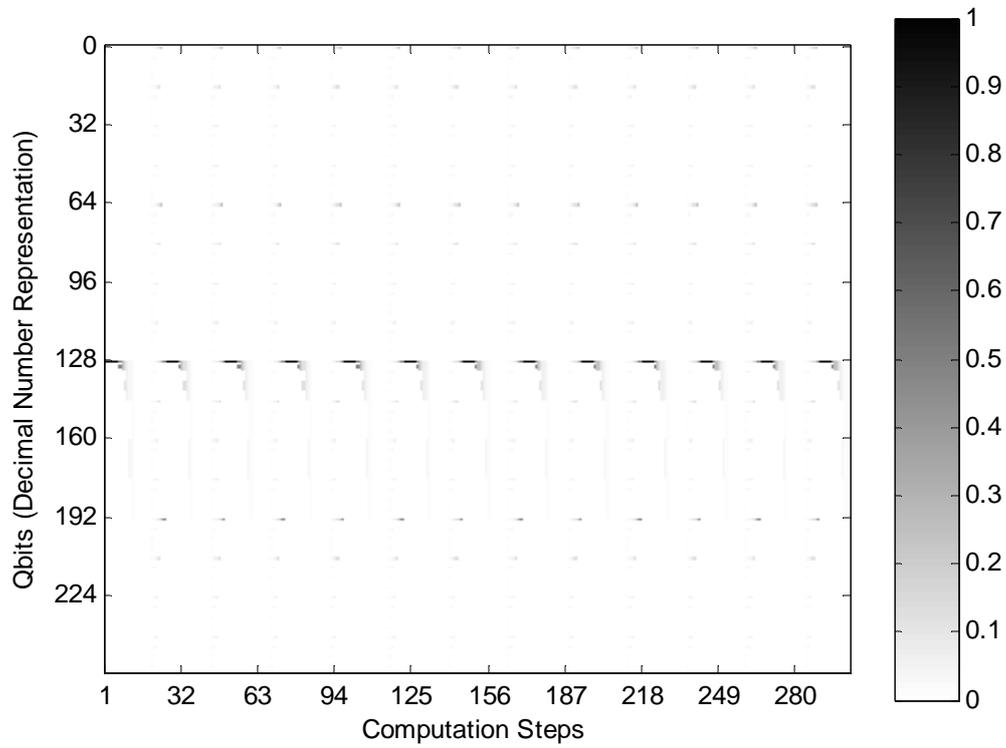

(b)

***FIG. 4**. Evolution of QCAs with four cells with Hadamard gates applied to both s-qubits and the c-qubits in each cell at evaluation phases. (a) The rule is the one shown in FIG. 1(b). (b) The rule is the one shown in FIG. 1(c).*

FIGs. 4 (a) and (b) show the simulated evolution of a QCA with four cells and initial state $|10000000\rangle$. The rules used for the evolution of these QCAs are the ones shown in FIGs. 1(b) and 1(c), respectively. At evaluation phases $H$ gates applied to both s-qubits and the c-qubits in each cell at evaluation phases. The periodic structure of the probability distribution patterns produced by the simulated evolution of all QCAs is apparent.

## IV. CONCLUSION

The evolution of the QCAs defined here is unitary and is linear because of the base-state superposition. Simulations provided a strong indication for periodic evolution. Furthermore, the period of the probability distribution patterns varies with the evolution rule and the number of cells. This is an indication for possible ability to construct QCAs with desirable periods of evolution. If this is possible, then many applications can be found for QCAs with two qubits per cell in quantum information processing.